\documentclass[eqsecnum,showpacs,preprint,amsmath,aps,nofootinbib]{revtex4}  %
\usepackage{epsfig}
\usepackage{amssymb}
\usepackage[dvips]{color}
\usepackage[latin1]{inputenc}
\usepackage{graphicx}

\def\beq{\begin{equation}}
\def\eeq{\end{equation}}
\def\beqq{\begin{eqnarray}}
\def\eeqq{\end{eqnarray}}

\newcommand{\bdm}{\begin{displaymath}}
\newcommand{\edm}{\end{displaymath}}

\def\pmb#1{\setbox0=\hbox{$#1$}%
  \kern-.025em\copy0\kern-\wd0
  \kern.05em\copy0\kern-\wd0
  \kern-.025em\raise.0433em\box0}

\makeatletter
\renewcommand*{\@fnsymbol}[1]{\ensuremath{\ifcase#1\or *\or \dagger\or
    \ddagger\or 
   \mathsection\or **\or \dagger\dagger
   \or \ddagger\ddagger \else\@ctrerr\fi}}
\makeatother

\begin{document}
\title{Restricted 3-body problem in effective-field-theory models
of gravity}

\author{Emmanuele Battista}
\email[E-mail: ]{ebattista@na.infn.it}
\affiliation{Dipartimento di Fisica, Complesso Universitario 
di Monte S. Angelo, Via Cintia Edificio 6, 80126 Napoli, Italy\\
INFN, Sezione di Napoli, Complesso Universitario di Monte
S. Angelo, Via Cintia Edificio 6, 80126 Napoli, Italy}

\author{Giampiero Esposito}
\email[E-mail: ]{gesposit@na.infn.it}
\affiliation{Istituto Nazionale di Fisica Nucleare, Sezione di
Napoli, Complesso Universitario di Monte S. Angelo, 
Via Cintia Edificio 6, 80126 Napoli, Italy}

\date{\today}

\begin{abstract}
One of the outstanding problems of classical celestial mechanics 
was the restricted $3$-body problem, in which a planetoid of small
mass is subject to the Newtonian attraction of two celestial bodies
of large mass, as it occurs, for example, in the sun-earth-moon
system. On the other hand, over the last decades, a systematic
investigation of quantum corrections to the Newtonian potential 
has been carried out in the literature on quantum gravity. 
The present paper studies
the effect of these tiny quantum corrections on the evaluation
of equilibrium points. It is shown that, despite the extreme
smallness of the corrections, there exists no choice of sign
of these corrections for which all qualitative features of the 
restricted $3$-body problem in Newtonian theory remain unaffected.
Moreover, first-order stability of equilibrium points is 
characterized by solving a pair of algebraic equations
of fifth degree, where some coefficients depend on the Planck length.
The coordinates of stable equilibrium points are slightly changed 
with respect to Newtonian theory, because the planetoid is no longer
at equal distance from the two bodies of large mass.
The effect is conceptually interesting but too small to be
observed, at least for the restricted $3$-body problems
available in the solar system.  
\end{abstract}


\maketitle

\section{Introduction}

It is frequently the case, in physics, that an hybrid scheme,
logically incomplete, turns out to be quite useful because the full
theory is unknown or leads to equations that cannot be solved.
Among the many conceivable examples of this feature, we mention
the following, since they are relevant for motivating the research
problem we are going to study.
\vskip 0.3cm
\noindent
(i) The nonrelativistic particle in curved spacetime 
\cite{DeWitt2003}, where the Schrodinger 
equation is studied, which is part
of nonrelativistic quantum theory, but the potential in such equation
receives a contribution from spacetime curvature, which is instead
defined and studied in general relativity.
\vskip 0.3cm
\noindent
(ii) Quantum field theory in curved spacetime, where the right-hand
side of the Einstein equations is replaced by the expectation value
of the regularized and renormalized energy-momentum tensor
$\langle T_{\mu \nu} \rangle$ evaluated in a classical spacetime
geometry. Only at a subsequent stage does one try to consider the
backreaction on the Einstein tensor, which, being coupled to a
nonclassical object like $\langle T_{\mu \nu} \rangle$, cannot remain
undisturbed.
\vskip 0.3cm
\noindent
(iii) The application of the effective field theory point of view
to the quantization of Einstein's general relativity. Within this
framework, starting from the Lagrangian density 
\begin{equation}
{\cal L} \equiv {\sqrt{-g}} \left[{c^{4}\over 16 \pi G} R
+{\cal L}_{{\rm matter}} \right],
\label{(1.1)}
\end{equation}
one includes all possible higher derivative couplings of the fields
in the gravitational Lagrangian. By doing so, any field singularities
generated by loop diagrams can be associated with some component of
the action and can be absorbed through a redefinition of the coupling
constants of the theory. By treating all coupling coefficients as
experimentally determined in this way, the effective field theory is
finite and singularity-free at any finite order of the loop
expansion \cite{D03}, even though it remains true that
Einstein's gravity is not perturbatively renormalizable 
\cite{DeWitt2003} and not even 2-loop on-shell finite \cite{Goroff1986}.
\vskip 0.3cm
\noindent
(iv) Among the many outstanding problems of classical physics and,
in particular, classical celestial mechanics, the 3-body problem
played a major role, and the genius of Poincar\'e himself 
\cite{Poincare1890} was not enough to arrive at a complete solution.
Nevertheless, one finds it often of interest, for example in the
analysis of the Sun-Earth-Moon system, to consider the so-called
restricted 3-body problem \cite{Pars}. In this case a body $A$
of mass $\alpha$ and a body $B$ of mass $\beta < \alpha$ move under
their mutual attraction. The center of mass $C$ of the $2$ bodies
moves uniformly in a straight line, and one can suppose it to be at
rest without loss of generality. The initial conditions tell us that
the orbit of $B$ relative to $A$ is a circle, hence the orbit of each
body relative to $C$ is a circle as well. Moreover, a third body, the
planetoid $P$, moves in the plane of motion of $A$ and $B$. By 
hypothesis, $P$ is subject to the Newtonian attraction of $A$ and
$B$, but its mass $m$ is so small that it cannot affect the motion
of $A$ and $B$. The problem consists therefore in evaluating the
motion of $P$.

Now when general relativity is viewed as an effective field theory,
it becomes of interest to derive (at least) the leading classical and
quantum corrections to the Newtonian potential of two large 
nonrelativistic masses. Hence we have been led to ask ourselves
whether, despite the extremely small numbers involved, a quantum
perspective on the restricted 3-body problem can be obtained.
The question is not merely of academic interest. Indeed, on the one
hand, we know already that very small quantities may produce nontrivial
effects in physics. An example, among the many, is provided by the
Stark effect: no matter how small is the external electric field,
the Stark-effect Hamiltonian has absolutely continuous spectrum on
the whole real line \cite{ReedSimon1978}, whereas the unperturbed
Hamiltonian for hydrogen atom has discrete spectrum on the negative
half-line. Yet another relevant example is provided by singular
perturbations in quantum mechanics: if a one-dimensional harmonic
oscillator is perturbed by a term proportional to negative powers
of the position operator, then no matter how small is the weight
coefficient one cannot recover the original Hamiltonian if the
perturbation is switched off. The unperturbed Hamiltonian has in 
fact both even and odd eigenfunctions, whereas the singular 
perturbation enforces the stationary states to vanish at the origin,
and the latter condition survives if the perturbation gets switched off
\cite{Klauder}, so that one eventually recovers a sort of
`halved' harmonic oscillator, with only half of the original
eigenfunctions.

On the other hand, by virtue of the improved technology with respect
to the golden age of Poincar\'e, it becomes conceivable to send off
satellites in the solar system that, within our lifetime, might become
part of suitable 3-body systems with the advantage, with respect to
natural planetoids such as the moon, that the satellite can be
`instructed' to approach and even nearly miss the large masses of
$A$ and $B$. Hence the putative quantum corrected Newtonian potential
can be tested at very small distances, in circumstances which were
inconceivable a century ago.

Section II builds the quantum-corrected Lagrangian of our model.
Section III writes down the equilibrium conditions and the 
partial derivatives of our full potential up to the second order.
Section IV is devoted to the equilibrium points on the line joining
$A$ to $B$, while Sec. V studies equilibrium points not lying
on the line that joins $A$ to $B$. Section VI identifies the 
unstable and stable equilibrium points. Concluding remarks and
open problems are presented in Sec. VII.  

\section{Quantum corrected Lagrangian of the model}

Following Ref. \cite{Pars} we take rotating axes with center of mass
$C$ as origin, and $CB$ as axis of $x$ (see Fig. 1). 
The length $AB$ is denoted
by $l$, and the angular velocity by $\omega$, so that
\begin{equation}
\omega^{2}={G(\alpha+\beta)\over l^{3}}.
\label{(2.1)}
\end{equation}
By doing so, we choose to neglect any correction, either classical
or quantum, to the Newtonian potential between the bodies having 
large mass. Thus, $A$ is permanently at rest, relative to the
rotating axes, at the point of coordinates $(-a,0)$, and $B$ is
permanently at rest at the point $(b,0)$, where \cite{Pars}
\begin{equation}
a={\beta \over (\alpha+\beta)}l, \;
b={\alpha \over (\alpha+\beta)}l.
\label{(2.2)}
\end{equation}

\begin{figure}
\includegraphics[scale=0.35]{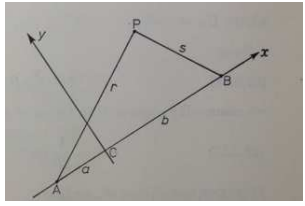}
\caption{The figure shows the two bodies of large mass,
$A$ and $B$, the center of mass $C$, and the planetoid at $P$.}
\end{figure}

The motion of the planetoid at $P(x,y)$ is the same as it would be
if $A$ and $B$ were constrained to move as they do, hence the
kinetic energy reads as
\begin{equation}
T={m \over 2}[({\dot x}-y \omega)^{2}
+({\dot y}+x \omega)^{2}].
\label{(2.3)}
\end{equation}
Furthermore, on denoting by $r$ the distance $AP$ and by $s$ the
distance $BP$, i.e.
\begin{equation}
r^{2}=(x+a)^{2}+y^{2}, \;
s^{2}=(x-b)^{2}+y^{2},
\label{(2.4)}
\end{equation}
the interaction potential is here taken to be
\begin{equation}
V=-{G m \alpha \over r}\left(1+{k_{1}\over r}
+{k_{2}\over r^{2}}\right)
-{Gm \beta \over s}\left(1+{k_{3}\over s}
+{k_{4}\over s^{2}}\right),
\label{(2.5)}
\end{equation}
where, on denoting by $\kappa_{1},\kappa_{2},\kappa_{3}$ three
dimensionless constants, one has
\begin{equation}
k_{1}=\kappa_{1}{G(m+\alpha)\over c^{2}},
\label{(2.6)}
\end{equation}
\begin{equation}
k_{2}=k_{4}=\kappa_{2}{G \hbar \over c^{3}}
=\kappa_{2} l_{P}^{2},
\label{(2.7)}
\end{equation}
\begin{equation}
k_{3}=\kappa_{3}{G(m+\beta)\over c^{2}}.
\label{(2.8)}
\end{equation}
In these formulas, $k_{1}$ and $k_{3}$ describe a classical
(post-Newtonian) contribution, whereas $k_{2}=k_{4}$ describes a
truly quantum correction. One arrives at these formulas through
a rather involved Feynman-diagram analysis, and the
$\kappa_{1},\kappa_{2},\kappa_{3}$ values obtained in Refs. 
\cite{D94,D03} differ both for the sign and their magnitude, 
because such References find 
\begin{equation}
\kappa_{1}=3 \; {\rm or} \; -1, 
\label{(2.9)}
\end{equation}
\begin{equation}
\kappa_{2}={41 \over 10 \pi} \; 
{\rm or} \; -{127 \over 30 \pi^{2}},
\label{(2.10)}
\end{equation}
respectively. In Ref. \cite{D94}, the author evaluated all corrections
resulting from vertex and vacuum polarization, whereas in Ref.
\cite{D03} the authors considered all diagrams for a scattering 
process. However, if one needs to iterate the lowest order potential
in some way, one should probably not include at least the box diagram.
Thus, the result in Ref. \cite{D03} is closer to the full answer, 
but it depends on some of the details of how one is going to use it.
We are grateful to the author of Ref. \cite{D94} for making all this
clear to us.

Our quantum corrected Lagrangian is therefore assumed 
to take the form
\begin{eqnarray}
{L \over m} &=& {1\over 2}({\dot x}^{2}+{\dot y}^{2})
+\omega (x{\dot y}-y{\dot x})+{1\over 2}\omega^{2}(x^{2}+y^{2})
\nonumber \\
&+& {G \alpha \over r}\left(1+{k_{1}\over r}+{k_{2}\over r^{2}}
\right)+{G \beta \over s}\left(1+{k_{3}\over s}
+{k_{2}\over s^{2}}\right) 
\nonumber \\
&=& T-V=T_{2}+T_{1}+T_{0}-V,
\label{(2.11)}
\end{eqnarray}
having denoted by $T_{n}$ the part of $T$ containing 
$n$-th order derivatives of $x$ or $y$. Such a Lagrangian does not
depend on $t$ explicitly, and the Jacobi integral \cite{Pars}
for it exists and is given by
\begin{equation}
J=T_{2}+V-T_{0},
\label{(2.12)}
\end{equation}
where, by virtue of (2.1) and (2.5),
\begin{equation}
T_{0}-V=G U,
\label{(2.13)}
\end{equation}
having set
\begin{equation}
U \equiv {1\over 2}{(\alpha+\beta)\over l^{3}}(x^{2}+y^{2})
+{\alpha \over r}\left(1+{k_{1}\over r}+{k_{2}\over r^{2}}
\right)+{\beta \over s}\left(1+{k_{3}\over s}
+{k_{2}\over s^{2}}\right).
\label{(2.14)}
\end{equation}
The resulting Lagrange equations of motion read as
\begin{equation}
{\ddot x}-2 \omega {\dot y}=G {\partial U \over \partial x},
\label{(2.15)}
\end{equation}
\begin{equation}
{\ddot y}+2 \omega {\dot x}=G {\partial U \over \partial y}.
\label{(2.16)}
\end{equation}
Since, from (2.12) and (2.13), $J=T_{2}-GU$, one has the simple
but nontrivial restriction according to which the motion of $P$
is only possible where
\begin{equation}
GU+J=T_{2}>0 \Longrightarrow U > -{J \over G}.
\label{(2.17)}
\end{equation}

\section{Equilibrium conditions and derivatives of the full potential}

The equilibrium points, either stable or unstable, are points at
which the full potential (2.14) is stationary, and hence one has
to study its first and second partial derivatives. To begin, one finds
\begin{equation}
{\partial U \over \partial x}=(\alpha+\beta){x \over l^{3}}
-{\alpha (x+a)\over r^{3}}\left(1+2{k_{1}\over r}
+3{k_{2}\over r^{2}}\right)
-{\beta(x-b)\over s^{3}}\left(1+2{k_{3}\over s}
+3{k_{2}\over s^{2}}\right).
\label{(3.1)}
\end{equation}
Thus, on using (2.2) and defining (cf. the classical formulas
in Ref. \cite{Pars})
\begin{equation}
\lambda \equiv {(\alpha+\beta)\over l^{3}}
-{\alpha \over r^{3}}\left(1+2{k_{1}\over r}+3{k_{2}\over r^{2}}
\right)-{\beta \over s^{3}}\left(1+2{k_{3}\over s}
+3{k_{2}\over s^{2}}\right),
\label{(3.2)}
\end{equation}
one can re-express ${\partial U \over \partial x}$ in the form
(see Fig. 2) 
\begin{equation}
{\partial U \over \partial x}=\lambda x 
+{\alpha \beta l \over (\alpha + \beta)}\left[{1\over s^{3}}
\left(1+2{k_{3}\over s}+3{k_{2}\over s^{2}}\right)
-{1\over r^{3}}\left(1+2{k_{1}\over r}+3{k_{2}\over r^{2}}
\right)\right],
\label{(3.3)}
\end{equation}
while, with the same notation, the other first derivative reads as
\begin{equation}
{\partial U \over \partial y}=\lambda y.
\label{(3.4)}
\end{equation}
For this to vanish, it is enough that either $y$ or $\lambda$ 
vanishes, in complete formal analogy with the classical
case \cite{Pars}. When $y=0$, the equilibrium points lie on the
line joining A to B, while the condition $\lambda=0$ yields
the equilibrium points not lying on the line joining A to B. 
Second derivatives of $U$ and their sign are important to understand
the nature of equilibrium points. For this purpose, we need the
first derivatives of the function $\lambda$, which are found to be
\begin{equation}
{\partial \lambda \over \partial x}={(x+a)\over r^{5}}\alpha
\left(3+8{k_{1}\over r}+15{k_{2}\over r^{2}}
\right)+{(x-b)\over s^{5}}\beta \left(3+8{k_{3}\over s}
+15{k_{2}\over s^{2}}\right),
\label{(3.5)}
\end{equation}
\begin{equation}
{\partial \lambda \over \partial y}=y \left[{\alpha \over r^{5}}
\left(3+8{k_{1}\over r}+15{k_{2}\over r^{2}}\right)
+{\beta \over s^{5}}\left(3+8{k_{3}\over s}
+15 {k_{2} \over s^{2}}\right)\right],
\label{(3.6)}
\end{equation}
by virtue of the identities (see (2.4))
\begin{equation}
{\partial r \over \partial x}={(x+a)\over r}, \; 
{\partial r \over \partial y}={y \over r}, \;
{\partial s \over \partial x}={(x-b)\over s}, \;
{\partial s \over \partial y}={y \over s}.
\label{(3.7)}
\end{equation}
The second derivatives of $U$ are hence given by
(see Figs. 3, 4 and 5)
\begin{equation}
{\partial^{2}U \over \partial x^{2}}=\lambda
+(x+a)^{2}{\alpha \over r^{5}}\left(3+8{k_{1}\over r}
+15 {k_{2}\over r^{2}}\right)
+(x-b)^{2}{\beta \over s^{5}}\left(3+8{k_{3}\over s}
+15 {k_{2}\over s^{2}}\right),
\label{(3.8)}
\end{equation}
\begin{equation}
{\partial^{2}U \over \partial x \partial y}
=y \left[{(x+a)\over r^{5}}\alpha \left(3+8{k_{1}\over r}
+15 {k_{2}\over r^{2}}\right)
+{(x-b)\over s^{5}}\beta \left(3+8{k_{3}\over s}
+15{k_{2}\over s^{2}}\right)\right],
\label{(3.9)}
\end{equation}
\begin{equation}
{\partial^{2}U \over \partial y^{2}}=\lambda+y^{2}\left[
{\alpha \over r^{5}}\left(3+8{k_{1}\over r}+15{k_{2}\over r^{2}}
\right)+{\beta \over s^{5}}\left(3+8{k_{3}\over s}
+15{k_{2}\over s^{2}}\right)\right].
\label{(3.10)}
\end{equation}
 
\begin{figure}
\includegraphics[scale=0.35]{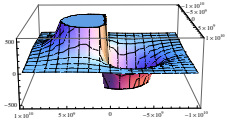}
\caption{Plot of the partial derivative with respect to the
$x$-coordinate of the potential $U(x,y)$ obtained by setting
$\lambda=0$. The graph has been obtained with the choice of
negative signs in (2.9) and (2.10) and for the system consisting
of Jupiter and two of its satellites, i.e. Adrastea and Ganymede.
For this system one has the following parameters:
$\alpha=m_{\rm Jupiter}=1.90 \times 10^{27}Kg$,
$\beta=m_{\rm Ganymede}=1.48 \times 10^{23}Kg$,
$m=m_{\rm Adrastea}=7.5 \times 10^{15}Kg$,
$l=1.07 \times 10^{9}$m, $a=8.33 \times 10^{5}$m,
$b=1.07 \times 10^{9}$m.}
\end{figure}

\begin{figure}
\includegraphics[scale=0.35]{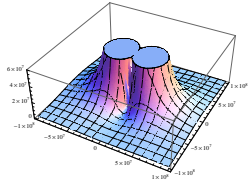}
\caption{Plot of the partial derivative $U_{,xx}$ obtained by
setting $\lambda=0$. The graph has been obtained with the choice
of positive signs in (2.9) and (2.10) and for the system consisting
of Sun, Earth and Moon. For this system one has the following
parameters: $\alpha=m_{\rm Sun}=1.99 \times 10^{30}Kg$,
$\beta=m_{\rm Earth}=5.97 \times 10^{24}Kg$,
$m=m_{\rm Moon}=7.35 \times 10^{22}Kg$,
$l=1.50 \times 10^{11}$m, 
$a=4.49 \times 10^{5}$m, $b=1.49 \times 10^{11}$m.}
\end{figure}

\begin{figure}
\includegraphics[scale=0.35]{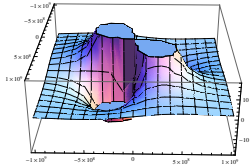}
\caption{Plot of the partial derivative $U_{,xy}$ obtained by
setting $\lambda=0$. The graph has been obtained with the choice
of positive signs in (2.9) and (2.10) and for the system consisting
of Jupiter and its satellites Adrastea and Ganymede.}
\end{figure}

\begin{figure}
\includegraphics[scale=0.35]{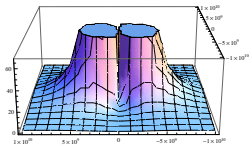}
\caption{Plot of the partial derivative $U_{,yy}$ obtained by
setting $\lambda=0$. The graph has been obtained with the choice
of negative signs in (2.9) and (2.10) and for the system consisting
of Sun, Earth and Moon.}
\end{figure}

\section{Equilibrium points on the line joining $A$ to $B$}

The line joining $A$ to $B$ is an axis having equation $y=0$, and
it can be divided into 3 regions (see Figs. 6, 7 and 8):
$$
{\cal R}_{1}: \; x \in ]-\infty,-a[, \;
{\cal R}_{2}: \; x \in ]-a,b[, \;
{\cal R}_{3}: \; x \in ]b,\infty[.
$$

\begin{figure}
\includegraphics[scale=0.35]{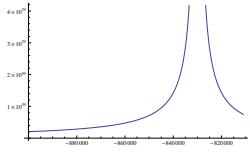}
\caption{Plot of the potential $U(x,0)$ in the region ${\cal R}_{1}$.
The graph has been obtained with the choice of positive signs in
(2.9) and (2.10) and for the system consisting of Jupiter and its
satellites Adrastea and Ganymede.}
\end{figure}

\begin{figure}
\includegraphics[scale=0.35]{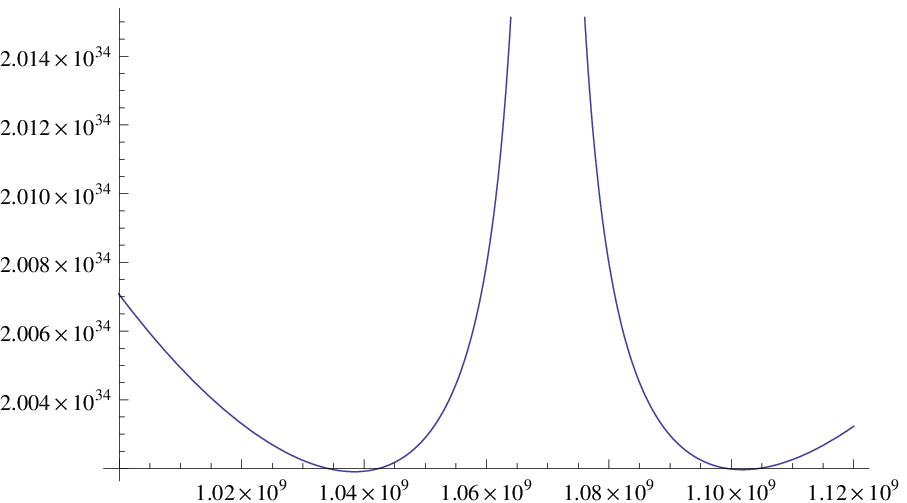}
\caption{Plot of the potential $U(x,0)$ in the region ${\cal R}_{2}$.
The graph has been obtained with the choice of positive signs in
(2.9) and (2.10) and for the system consisting of Jupiter and its
satellites Adrastea and Ganymede.}
\end{figure}

\begin{figure}
\includegraphics[scale=0.35]{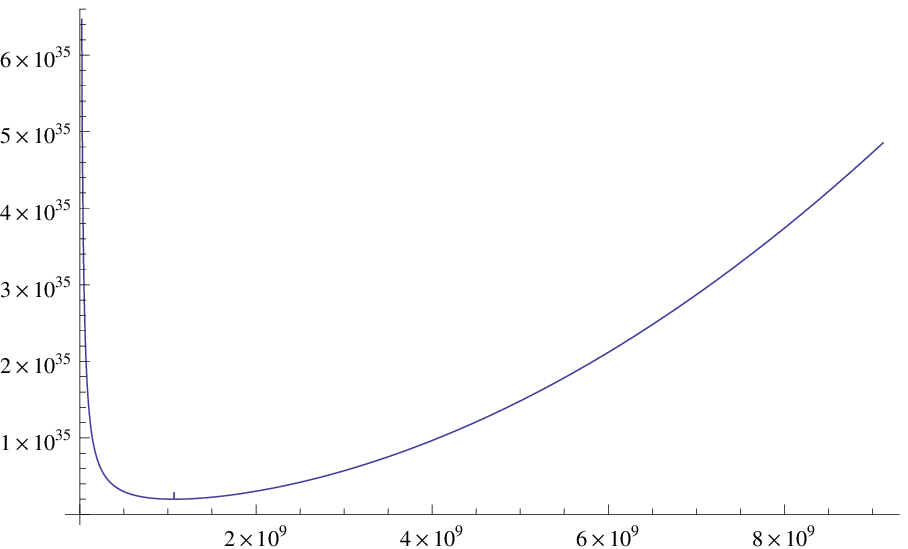}
\caption{Plot of the potential $U(x,0)$ in the region ${\cal R}_{3}$.
The graph has been obtained with the choice of positive signs in
(2.9) and (2.10) and for the system consisting of Jupiter and its
satellites Adrastea and Ganymede.}
\end{figure}

From Eq. (2.4) and $y=0$ one has 
$r=|x+a|,s=|x-b|$, and hence Eqs. (3.2) and (3.8) yield
\begin{equation}
\left . {\partial^{2}U \over \partial x^{2}} \right |_{y=0}
=\left[{(\alpha+\beta)\over l^{3}}+2{\alpha \over r^{3}}
+2{\beta \over s^{3}}\right]
+2{\alpha \over r^{4}}\left(3 k_{1}+6{k_{2}\over r}\right)
+2{\beta \over s^{4}}\left(3k_{3}+6{k_{2}\over s}\right).
\label{(4.1)}
\end{equation}
In Newtonian theory, since all terms in square brackets in (4.1)
are positive, one concludes that $U_{,xx}$ is always positive on
$y=0$. However, by virtue of (2.5)-(2.10), this may no longer be true
in our case, if one adopts the negative signs on the right-hand 
side of (2.9) and (2.10) and if one lets either $r$ or $s$ or both
to approach $0$. Thus, the sufficient condition for preservation of the
sign in Newtonian theory reads as
\begin{equation}
\boxed{
\left(3k_{1}+6{k_{2}\over r}\right)+{\beta \over \alpha}
\left({r \over s}\right)^{4}\left(3k_{3}+6{k_{2}\over s}\right)>0},
\label{(4.2)}
\end{equation}
which is however violated with the choice of negative signs in (2.9)
and (2.10).

Note that the function $U(x,0)$ has, from (2.14), the limiting
behavior 
\begin{equation}
\lim_{x \to -a}U(x,0)=\lim_{x \to b}U(x,0)=+\infty,
\label{(4.3)}
\end{equation}
\begin{equation}
\lim_{x \to -\infty}U(x,0)=\lim_{x \to +\infty}U(x,0)=+\infty.
\label{(4.4)}
\end{equation}
Moreover, $U_{,x}$ passes just once through $0$ 
in each of the $3$ regions ${\cal R}_{1},{\cal R}_{2}$ and 
${\cal R}_{3}$, which implies that there exist $3$ equilibrium
points on $AB$, when $U$ has minima at the points 
$N_{1}(x=n_{1}),N_{2}(x=n_{2})$ and $N_{3}(x=n_{3})$.

To study the location of the equilibrium points, we note,
following Ref. \cite{Pars}, that
\begin{equation}
{r\over (x+a)}=(-1,1,1), \;
{s \over (x-b)}=(-1,-1,1),
\label{(4.5)}
\end{equation}
the $3$ values on the right-hand side referring to 
${\cal R}_{1},{\cal R}_{2}$ and ${\cal R}_{3}$, respectively,
so that in ${\cal R}_{1}$ for example (see (3.3))
\begin{equation}
{\partial U \over \partial x}=(\alpha+\beta){x \over l^{3}}
+{\alpha \over r^{2}}\left(1+2{k_{1}\over r}+3{k_{2}\over r^{2}}
\right)+{\beta \over s^{2}}\left(1+2{k_{3}\over s}
+3{k_{2}\over s^{2}}\right).
\label{(4.6)}
\end{equation}
At the point $x=-a-l$ one has $r=l,s=2l$, and from (2.2) and 
(4.6) one finds
\begin{equation}
\left . {\partial U \over \partial x}\right |_{x=-a-l}
=-{7\over 4}{\beta \over l^{2}}
+{1\over l^{3}}\left[\alpha \left(2k_{1}+3{k_{2}\over l}\right)
+{\beta \over 4}\left(k_{3}+{3\over 4}{k_{2}\over l}\right)
\right].
\label{(4.7)}
\end{equation}
In Newtonian theory, the sum in square brackets in (4.7) is
absent and one can say that $U_{,x}$ is negative and hence
$N_{1}$ lies between $x=-a-l$ and $x=-a$. In our model, for this
to remain true, one should impose the sufficient condition
\begin{equation}
\boxed{
2k_{1}+3{k_{2}\over l}+{\beta \over 4 \alpha}\left(k_{3}
+{3\over 4}{k_{2}\over l}\right) <0},
\label{(4.8)}
\end{equation}
which is however violated with the choice of positive signs in
(2.9) and (2.10).

Similarly, to understand whether the equilibrium point $N_{2}$
lies between $C$ and $B$, one has to evaluate $U_{,x}$ at $C$,
where $r=a,s=b,x=y=0$, which yields, from (3.3),
\begin{equation}
\left . {\partial U \over \partial x} \right |_{C}
=-(\alpha^{3}-\beta^{3}){(\alpha+\beta)^{2}\over 
\alpha^{2}\beta^{2}l^{2}}
-\left[{\alpha \over a^{3}}\left(2k_{1}+3{k_{2}\over a}
\right)+{\beta \over b^{3}}\left(2k_{3}+3{k_{2}\over b}\right)
\right].
\label{(4.9)}
\end{equation}
In Newtonian theory, the sum in square brackets in (4.9) does
not occur, and hence $\left . {\partial U \over \partial x}
\right |_{C}$ is always negative. For this to remain true in our
model, one has to impose the sufficient condition
\begin{equation}
\boxed{
k_{1}+{3\over 2}{k_{2}\over a}+{\beta \over \alpha}
\left({a\over b}\right)^{3}\left(k_{3}+{3\over 2}
{k_{2}\over b}\right) >0},
\label{(4.10)}
\end{equation}
which is instead violated with the choice of negative signs
in (2.9) and (2.10).

At this stage, despite the incompleteness of our analysis, we
have already proved a simple but nontrivial result: 
{\it not only can our model be used to discriminate among competing
theories of effective gravity, but there exists no choice of signs
in (2.9) and (2.10) for which all qualitative features of the
restricted $3$-body problem in Newtonian theory remain unaffected.
As far as we can see, this means that either we reject effective
theories of gravity or we should expect them to be able to lead to
testable effects in suitable $3$-body systems, e.g. a satellite
which is programmed to approach very closely (much closer than
the moon can afford approaching the earth) 
$2$ celestial bodies of large mass}.

Furthermore, from (3.10) we find
\begin{eqnarray}
\; & \; & \left . {\partial^{2}U \over \partial y^{2}}
\right |_{N_{1}}=\lambda={\alpha \beta l \over (\alpha+\beta)}
{1\over x}\left({1\over r^{3}}-{1\over s^{3}}\right)
\nonumber \\
&+& {1\over x}\left[2 \left({k_{1}\over r^{4}}-{k_{3}\over s^{4}}
\right)+3k_{2}\left({1\over r^{5}}-{1\over s^{5}}\right)\right].
\label{(4.11)}
\end{eqnarray}
In Newtonian theory, the sum of terms in square brackets in (4.11)
does not occur, and hence one points out that, since at $N_{1}$
$x$ is negative and $r<s$, the second derivative of $U$ at
$N_{1}$ is negative \cite{Pars}. In our model, however, the
sufficient condition for this to remain true, i.e.
\begin{equation}
\boxed{
\left({k_{1}\over r^{4}}-{k_{3}\over s^{4}}\right)
+{3\over 2}k_{2}\left({1\over r^{5}}-{1\over s^{5}} \right)>0},
\label{(4.12)}
\end{equation}
can be violated, for example, as $r \rightarrow 0$ with the
negative choice of sign in (2.10).

We note also that at $N_{2}$, where $r=x+a$ and $s=x-b$, one has 
from (3.10)
\begin{equation}
\left . {\partial^{2}U \over \partial y^{2}} \right |_{N_{2}}
={(\alpha +\beta)\over l^{3}}-{\alpha \over r^{3}}
-{\beta \over s^{3}}
-\left[2 \left(\alpha {k_{1}\over r^{4}}
+\beta {k_{3}\over s^{4}}\right)
+3k_{2}\left({\alpha \over r^{5}}+{\beta \over s^{5}}
\right)\right].
\label{(4.13)}
\end{equation}
In Newtonian theory, the sum of terms in square brackets in (4.13)
does not occur, and one finds that $U_{,yy}$ is negative at 
$N_{2}$, because in ${\cal R}_{2}$ both $r$ and $s$ are less than
$l$. In our model, for this to remain true, the following
sufficient condition should hold:
\begin{equation}
\boxed{
\alpha{k_{1}\over r^{4}}+\beta{k_{3}\over s^{4}}
+{3\over 2}k_{2}\left({\alpha \over r^{5}}
+{\beta \over s^{5}}\right)>0},
\label{(4.14)}
\end{equation}
which is however violated if the negative signs are chosen in
(2.9) and (2.10).

On reverting now to the graph of $U(x,0)$, there are minima
at $N_{1},N_{2}$ and $N_{3}$, and we would like to determine
at which of these $3$ points $U(x,0)$ has the greatest value,
and at which it has instead the least value. In Newtonian theory, 
one finds that $U(n_{2})>U(n_{3})>U(n_{1})$. To establish the
counterpart in our model, let $Q_{3}(x=q_{3})$ be the point
of ${\cal R}_{3}$ whose distance from $B$ is equal to the distance
of $N_{2}$ from $B$, i.e. $N_{2}B=BQ_{3}=j$. Thus, following 
patiently a number of cancellations, we find
\begin{eqnarray}
\; & \; & U(n_{2})-U(q_{3})=U(x=b-j,y=0,r=l-j,s=j)
\nonumber \\
&-& U(x=b+j,y=0,r=l+j,s=j)
\nonumber \\ 
&=& 2\alpha j \left({1\over (l-j)^{2}}-{1\over l^{2}}\right)
\nonumber \\
&+& {2 \alpha j \over (l-j)^{2}(l+j)^{2}}
\left[2k_{1}l +{k_{2}(j^{2}+3l^{2})\over (l^{2}-j^{2})}
\right].
\label{(4.15)}
\end{eqnarray}
In Newtonian theory, the sum of terms in square brackets in
(4.15) does not occur, and one therefore finds 
$U(n_{2})-U(q_{3})>0$. In our model, for this to remain true,
one should impose the following sufficient condition
\begin{equation}
\boxed{
k_{1}+{1\over 2}k_{2}{(j^{2}+3l^{2})\over l(l^{2}-j^{2})}>0},
\label{(4.16)}
\end{equation}
which is instead violated if the negative signs are chosen in 
(2.9) and (2.10).

Last, let $Q_{1}(x=q_{1})$ be the point of ${\cal R}_{1}$ whose
distance from $C$ is equal to the distance of $N_{3}$ from $C$,
i.e. $Q_{1}C=CN_{3}=f$. Then we find
\begin{eqnarray}
\; & \; & U(n_{3})-U(q_{1})=U(x=f,y=0,r=x+a,s=x-b)
\nonumber \\
&-& U(x=-f,y=0,r=x-a,s=-x+b) 
\nonumber \\
&=& {2\alpha \beta l (b^{2}-a^{2})\over (\alpha+\beta)
(f^{2}-a^{2})(f^{2}-b^{2})}
\nonumber \\
&+& {2\alpha \beta l \over (\alpha+\beta)(f^{2}-a^{2})^{2}
(f^{2}-b^{2})^{2}}
\biggr \{ 2f \Bigr[k_{3}(f^{2}-a^{2})^{2}
-k_{1}(f^{2}-b^{2})^{2}\Bigr]
\nonumber \\
&+& {k_{2}\Bigr[(b^{2}+3f^{2})(f^{2}-a^{2})^{3}
-(a^{2}+3f^{2})(f^{2}-b^{2})^{3}\Bigr]
\over (f^{2}-a^{2})(f^{2}-b^{2})} \biggr \}.
\label{(4.17)}
\end{eqnarray}
In Newtonian theory, the sum of terms in curly brackets in (4.17)
does not occur, and one finds $U(n_{3})>U(q_{1})$. In our model, for
this to remain true, one should impose the sufficient condition
\begin{equation}
\boxed{
2f \Bigr[k_{3}(f^{2}-a^{2})^{2}-k_{1}(f^{2}-b^{2})^{2}\Bigr]
+{k_{2}\Bigr[(b^{2}+3f^{2})(f^{2}-a^{2})^{3}
-(a^{2}+3f^{2})(f^{2}-b^{2})^{3}\Bigr] \over
(f^{2}-a^{2})(f^{2}-b^{2})}>0}.
\label{(4.18)}
\end{equation}
This is more involved than (4.16), and it is not a priori so
obvious whether a choice of signs in (2.9) and (2.10) leads
always to its fulfillment.
  
\section{Equilibrium points not lying on the line that joins
$A$ to $B$}

When the equilibrium points do not lie on the line joining $A$ to
$B$, the coordinate $y$ does not vanish and hence the first derivative
(3.4) vanishes because $\lambda=0$. On the other hand, the first
derivative (3.3) should vanish as well, which then implies,
by virtue of $\lambda=0$,
\begin{equation}
{1\over r^{3}}\left(1+2{k_{1}\over r}+3{k_{2}\over r^{2}}\right)
={1\over s^{3}}\left(1+2{k_{3}\over s}+3{k_{2}\over s^{2}}
\right).
\label{(5.1)}
\end{equation}
Unlike Newtonian theory \cite{Pars}, this equation is no longer
solved by $r=s$. The definition (3.2), jointly with (5.1),
makes it now possible to express the condition $\lambda=0$ in
the form
\begin{equation}
{1\over l^{3}}={1\over r^{3}}+2{k_{1}\over r^{4}}
+3{k_{2}\over r^{5}}.
\label{(5.2)}
\end{equation}
This is an algebraic equation of fifth degree in the variable
\begin{equation}
w \equiv {1\over r},
\label{(5.3)}
\end{equation}
and we divide both sides by $3k_{2}$ and exploit the definitions
(2.6)-(2.8) to write it in the form
\begin{equation}
\sum_{k=0}^{5}\zeta_{k}w^{k}=0, 
\label{(5.4)}
\end{equation}
where
\begin{equation}
\zeta_{5} \equiv 1,
\label{(5.5)}
\end{equation}
\begin{equation}
\zeta_{4} \equiv {2\over 3}{\kappa_{1}\over \kappa_{2}}
{G(m+\alpha)\over c^{2}l_{P}^{2}},
\label{(5.6)}
\end{equation}
\begin{equation}
\zeta_{3} \equiv {1\over 3 \kappa_{2}}{1\over l_{P}^{2}},
\label{(5.7)}
\end{equation}
\begin{equation}
\zeta_{2}=\zeta_{1} \equiv 0,
\label{(5.8)}
\end{equation}
\begin{equation}
\zeta_{0} \equiv -{1\over 3 \kappa_{2}}{1\over l_{P}^{2}l^{3}}.
\label{(5.9)}
\end{equation}
Since this equation is of odd degree with real coefficients, the
fundamental theorem of algebra guarantees the existence of at
least a real solution, despite the lack of a general solution
algorithm for all algebraic equations of degree greater than $4$.
Moreover, by virtue of the small term ${G\over c^{2}}$, the
coefficient $\zeta_{4}$ plays a negligible role both in the
sun-earth-moon system, where $\alpha=m_{{\rm sun}},
\beta=m_{{\rm earth}},m=m_{{\rm moon}},l=l_{{\rm sun-earth}}$, 
and in many other conceivable toy models of the restricted 
$3$-body problem, as is confirmed by detailed numerical checks.
We find only one positive root $w_{+}(l)$ of Eq. (5.4)
when the positive signs are chosen in (2.9) and (2.10),
following \cite{D03} (whereas $2$ positive roots are obtained
when negative signs are taken in (2.9) and (2.10)), from 
which $r(l)={1\over w_{+}(l)}$. Eventually, one can 
evaluate $s(l)=s(r(l))$ from
Eq. (5.1), which can be viewed as an algebraic equation of
fifth degree in the variable 
\begin{equation}
u \equiv {1\over s},
\label{(5.10)}
\end{equation}
i.e. (cf. Eq. (5.4))
\begin{equation}
\sum_{k=0}^{5}{\widetilde \zeta}_{k}u^{k}=0,
\label{(5.11)}
\end{equation}
where
\begin{equation}
{\widetilde \zeta}_{k}=\zeta_{k} \; \forall k=0,1,2,3,5,
\label{(5.12)}
\end{equation}
\begin{equation}
{\widetilde \zeta}_{4} \equiv {2\over 3}
{\kappa_{3}\over \kappa_{2}}
{G(m+\beta)\over c^{2}l_{P}^{2}}.
\label{(5.13)}
\end{equation}
Also in the case of Eq. (5.11) we have found only a positive
solution $u_{+}(l)$ both for the sun-earth-moon system and for 
any conceivable toy model for this restricted $3$-body problem.

The Cartesian coordinates $(x,y)$ of the equilibrium points not lying
along $AB$ can be found from the general formulas (2.4),
with the notation 
\begin{equation}
r(l) \equiv {1\over w_{+}(l)}, \;
s(l) \equiv {1\over u_{+}(l)},
\label{(5.14)}
\end{equation}
i.e.
\begin{equation}
r^{2}(l)=x^{2}+y^{2}+2ax+a^{2},
\label{(5.15)}
\end{equation}
\begin{equation}
s^{2}(l)=x^{2}+y^{2}-2bx+b^{2}.
\label{(5.16)}
\end{equation}
Subtraction of Eq. (5.16) from Eq. (5.15) yields
\begin{equation}
x(l) \equiv {(r^{2}(l)-s^{2}(l)+b^{2}-a^{2})\over 2(a+b)},
\label{(5.17)}
\end{equation}
while $y(l)$ can be obtained from (5.15) in the form
\begin{equation}
y_{\pm}(l) \equiv \pm \sqrt{r^{2}(l)-x^{2}(l)-2ax(l)-a^{2}}.
\label{(5.18)}
\end{equation}
Thus, there exist $2$ equilibrium points not lying on the line joining
$A$ to $B$, hereafter written in the form
\begin{equation}
N_{4}(x(l),y_{+}(l)), \; 
N_{5}(x(l),y_{-}(l)).
\label{(5.19)}
\end{equation}
In Newtonian theory, where $r=s$, the formula (5.19) reduces to the 
familiar \cite{Pars}
\begin{equation}
N_{4}\left({(\alpha-\beta)\over (\alpha+\beta)}{l\over 2},
{\sqrt{3}\over 2}l \right), \;
N_{5}\left({(\alpha-\beta)\over (\alpha+\beta)}{l\over 2},
-{\sqrt{3}\over 2}l \right), 
\label{(5.20)}
\end{equation}
by virtue of (2.2). The geometric interpretation of these formulas
is simple but it has a nontrivial consequence: 
at the points $N_{4}$ and $N_{5}$
the planetoid is not at the same distance from $A$ and $B$, unlike
Newtonian theory. Our quantum corrected model predicts a very tiny
displacement from the case $r=s$, but its effect cannot be
observed in the solar system, because in the available 
implementations of the restricted $3$-body problem the differences
\begin{equation}
\delta_{1}(l) \equiv x(l)-{(\alpha-\beta)\over (\alpha+\beta)}
{l\over 2}, \;
\delta_{2}(l) \equiv y_{+}(l)-{\sqrt{3}\over 2}l, \;
\delta_{3}(l) \equiv y_{-}(l)+{\sqrt{3}\over 2}l
\label{(5.21)}
\end{equation}
are too small to be observed, as is unfortunately the case for
many interesting effects in quantum gravity.
 
\section{Unstable and stable equilibrium points}

A rather important question is whether the positions of equilibrium
are stable. In the affirmative case, the planetoid would therefore
remain permanently near the point of stable equilibrium.
To study this issue, on denoting by $(x_{0},y_{0})$ one of the points
$N_{1},N_{2},N_{3},N_{4},N_{5}$, one writes in the equations of
motion (2.15) and (2.16)
\begin{equation}
x=x_{0}+\xi, \; y=y_{0}+\eta.
\label{(6.1)}
\end{equation}
By expanding the right-hand sides in powers of $\xi$ and $\eta$, and
retaining only terms of first order, one obtains the linear
approximation \cite{Pars}
\begin{equation}
{\ddot \xi} -2 \omega {\dot \eta}=G(A\xi + B \eta),
\label{(6.2)}
\end{equation}
\begin{equation}
{\ddot \eta} +2 \omega {\dot \xi}=G(B\xi + C \eta),
\label{(6.3)}
\end{equation}
having defined
\begin{equation}
A \equiv \left . {\partial^{2}U \over \partial x^{2}} 
\right |_{x_{0},y_{0}} , \;
B \equiv \left . {\partial^{2}U \over \partial x \partial y}
\right |_{x_{0},y_{0}} , \;
C \equiv \left . {\partial^{2}U \over \partial y^{2}} 
\right |_{x_{0},y_{0}}.
\label{(6.4)}
\end{equation}
Equations (6.2) and (6.3) are a coupled set of ordinary differential
equations with constant coefficients, and hence one can look for
its solution in the form
\begin{equation}
\xi=\xi_{0}{\rm e}^{{t \over \tau}}, \; 
\eta=\eta_{0}{\rm e}^{{t \over \tau}}.
\label{(6.5)}
\end{equation}
This leads to the linear homogeneous system of algebraic equations
\begin{equation}
\left({1\over \tau^{2}}-GA \right)\xi
- \left(2{\omega \over \tau}+GB \right)\eta=0,
\label{(6.6)}
\end{equation}
\begin{equation}
\left(2{\omega \over \tau}-GB \right)\xi
+\left({1\over \tau^{2}}-GC \right)\eta=0.
\label{(6.7)}
\end{equation}
Nontrivial solutions exist if and only if the determinant of the
matrix of coefficients vanishes. Such a condition is expressed
by the algebraic equation of fourth degree
\begin{equation}
{1\over \tau^{4}}-[G(A+C)-4 \omega^{2}]{1\over \tau^{2}}
+G^{2}(AC-B^{2})=0.
\label{(6.8)}
\end{equation}
The variable is of course the square of 
${1\over \tau}$, and for it one
finds, from the standard theory of algebraic equations of
second degree,
\begin{equation}
{1\over \tau^{2}}={1\over 2}[G(A+C)-4 \omega^{2}] 
\pm {1\over 2}\sqrt{(G(A+C)-4 \omega^{2})^{2}
-4G^{2}(AC-B^{2})}.
\label{(6.9)}
\end{equation}

\subsection{Conditions for first-order instability 
of $N_{1},N_{2},N_{3}$}

In Newtonian theory, $(AC-B^{2})$ is negative at $N_{1},N_{2},N_{3}$,
and hence only half of the ${1\over \tau^{2}}$ 
values are negative, which
implies that the criterion for first-order stability \cite{Pars}
is not satisfied. In our model, it remains true, from (3.9), that
our $B$ vanishes at $N_{1},N_{2},N_{3}$, and we express our $A$ 
at $N_{1},N_{2},N_{3}$ from (4.1), 
our $C$ at $N_{1},N_{3}$ from (4.11),
and our $C$ at $N_{2}$ from (4.13). Thus, provided that the sufficient
conditions (4.2), (4.12) and (4.14) hold, which are in turn guaranteed,
as we know, from the choice of positive signs in (2.9) and (2.10),
it is always true that $(AC-B^{2})<0$, and the points 
$N_{1},N_{2},N_{3}$ remain points of unstable equilibrium
even in the presence of quantum corrections obtained from
an effective-gravity picture \cite{D03}.

\subsection{Conditions for first-order stability of $N_{4},N_{5}$}

At the points $N_{4}$ and $N_{5}$, the vanishing of $\lambda$
simplifies the evaluation of $A$ and $C$ from (3.8) and (3.10),
and we find (with the understanding that $r=r(l)$,  
$s=s(l)$ and $y=y(l)$ as in Sec. V)
\begin{equation}
A={\alpha(r^{2}-y^{2})\over r^{5}}
\left(3+8{k_{1}\over r}+15{k_{2}\over r^{2}}\right)
+{\beta(s^{2}-y^{2})\over s^{5}}
\left(3+8{k_{3}\over s}+15{k_{2}\over s^{2}}\right),
\end{equation}
\begin{equation}
C={\alpha y^{2}\over r^{5}}
\left(3+8{k_{1}\over r}+15{k_{2}\over r^{2}}\right)
+{\beta y^{2}\over s^{5}}
\left(3+8{k_{3}\over s}+15{k_{2}\over s^{2}}\right),
\end{equation}
\begin{eqnarray}
B^{2}&=& {\alpha^{2}y^{2}(r^{2}-y^{2})\over r^{10}}
\left(3+8{k_{1}\over r}+15{k_{2}\over r^{2}}\right)^{2}
+{\beta^{2}y^{2}(s^{2}-y^{2})\over s^{10}}
\left(3+8{k_{3}\over s}+15{k_{2}\over s^{2}}\right)^{2}
\nonumber \\
&+& {2\alpha \beta y^{2}\over r^{5}s^{5}}
\left(3+8{k_{1}\over r}+15{k_{2}\over r^{2}}\right)
\left(3+8{k_{3}\over s}+15{k_{2}\over s^{2}}\right)
(x^{2}+(a-b)x-ab).
\label{(6.10)}
\end{eqnarray}
In the evaluation of $(AC-B^{2})$ we find therefore exact
cancellation of the $2$ pairs of terms involving $\alpha^{2}$ 
and $\beta^{2}$. Moreover, on exploiting from (2.4) the identity
\begin{equation}
r^{2}+s^{2}=2(x^{2}+y^{2})+2(a-b)x+a^{2}+b^{2},
\label{(6.11)}
\end{equation}
we obtain, bearing in mind that $(a+b)=l$,
\begin{equation}
(AC-B^{2})={\alpha \beta y^{2}l^{2}\over r^{5}s^{5}}
\left(3+8{k_{1}\over r}+15{k_{2}\over r^{2}}\right)
\left(3+8{k_{3}\over s}+15{k_{2}\over s^{2}}\right).
\label{(6.12)}
\end{equation}
This is all we need, because it is clearly positive if
the positive signs are chosen in (2.9) and (2.10),
and it ensures that all values of ${1\over \tau^{2}}$ 
from the solution formula
(6.9) are negative (a result further confirmed by numerical
analysis for the sun-earth-moon and Jupiter-Adrastea-Ganymede
systems), in full agreement with the criterion for 
first-order stability \cite{Pars} of the equilibrium points.

\section{Concluding remarks and open problems}

Not only has the (restricted) $3$-body problem played an important role
in the historical development of celestial mechanics
\cite{Poincare1890, Poincare1892} and classical dynamics
\cite{Pars}, but it has also found important applications
to modern physics. For example, in Ref. \cite{Bial94}, the authors
have discovered, by analytic and numerical methods, the existence of
stable, although nonstationary, quantum states of electrons moving
on circular orbits that are trapped in an effective potential well
made of the Coulomb potential and the rotating electric field
produced by a strong circularly polarized electromagnetic wave.

In the theory of gravitation, the undisputable smallness of classical
and quantum corrections to the Newtonian potential had always
discouraged the investigation of their role in the restricted
$3$-body problem. Our contribution has been precisely a systematic
investigation of the ultimate consequences of such additional terms.
Our sufficient conditions (4.2), (4.8), (4.10), (4.12), (4.14),
(4.16) and (4.18) are original and imply that some changes of
qualitative features are unavoidable with respect to Newtonian
theory, regardless of the choice of signs made in (2.9) and (2.10),
although $6$ out of $7$ sufficient conditions are fulfilled
with the choice of positive signs in (2.9) and (2.10).
Section V has shown that the equilibrium points not lying on the
line that joins $A$ to $B$ are found by solving a pair of algebraic
equations of fifth degree, and their coordinates have been
obtained for the first time in the class of effective theories 
of gravity studied in Refs. \cite{D03,D94}. Section VI has studied  
first-order stability for the $5$ equilibrium points 
of the problem. We have proved therein that, provided the positive
signs are chosen in (2.9) and (2.10), the $3$ points along the line
joining $A$ to $B$ are unstable, while the $2$ points not on
$AB$ are stable equilibrium points to first order.

It now remains to be seen whether the present techniques in space
sciences make it possible to realize a satellite $P$ that approaches
so closely the celestial bodies $A$ and $B$ that our tiny
corrections start making themselves manifest. 
Unfortunately, the differences in (5.21) between quantum corrected
and Newtonian values of the coordinates of stable-equilibrium points
$N_{4}$ and $N_{5}$ are too small to be observed, at least in the
solar system. However, one cannot yet rule out that future 
technological developments will make it possible to  
ckeck against observations the
current effective theories of gravity, which would bring quantum
gravity research much closer to the experimental world. Last, but 
not least, the whole analysis performed in Refs. 
\cite{Poincare1890, Poincare1892}, if generalized to the extended
theories of gravity inspired by the works in Refs. 
\cite{D94,D03}, might lead to the discovery of novel features
of orbital motion.

\acknowledgments 
The authors are indebted to John Donoghue for enlightening 
correspondence. G. E. is grateful to the Dipartimento di Fisica of
Federico II University, Naples, for hospitality and support.

\end{document}